\documentclass[aps,pre,twocolumn]{revtex4-1}
\usepackage{amssymb}
\usepackage{amsmath,bm}
\usepackage{graphicx,color}
\usepackage{bbm}
\usepackage{multirow}
\newcommand{\B}[1]{{\bm{#1}}}

\begin{document}

\title{Instabilities of Time-averaged Configurations in Thermal Glasses}

\author{Prasenjit Das$^1$, Valery Ilyin$^1$ and Itamar Procaccia$^{1,2}$}
\affiliation{$^1$Department of Chemical Physics, the Weizmann Institute of Science, Rehovot 76100, Israel, 
	\\$^2$ Center for OPTical IMagery Analysis and Learning, Northwestern Polytechnical University, Xi'an, 710072 China}

\begin{abstract}
In amorphous solids at finite temperatures the particles follow chaotic trajectories which, at temperatures
sufficiently lower than the glass transition, are trapped in ``cages". Averaging their positions for times shorter
than the diffusion time, one can define a time-averaged configuration. Under strain or stress,
 these {\em average} configurations undergo sharp plastic instabilities. In athermal glasses the
 understanding of plastic instabilities is furnished by
 the Hessian matrix, its eigenvalues and eigenfunctions. Here we propose an uplifting
 of Hessian methods to thermal glasses, with the aim of understanding
 the plastic responses in the time averaged configuration. We discuss a number of potential
 definitions of Hessians and identify which of these can provide eigenvalues and eigenfunctions which
 can explain and predict the instabilities of the time-averaged configurations.
 The conclusion is that the non-affine changes in the
 average configurations during an instability is accurately predicted by the eigenfunctions of the low-lying
 eigenvalues of the chosen Hessian.

\end{abstract}

\maketitle

\section{Introduction}
The theory of the micro-mechanics of plasticity and material failure of amorphous solids at
finite temperatures lags behind the understanding of the athermal counterpart. At $T=0$ an amorphous
solids in mechanical equilibrium can be usefully characterized by the eigenfunctions and eigenvalues
of the Hessian matrix. In a system with $N$ particles in a volume $V$ in $d$ dimensions
and a Hamiltonian $U(\B r_1,\B r_2\dots,\B r_N)$, where the particles' coordinates
are $\{\B r_i\}_{i=1}^N$, the Hessian matrix is defined as
\begin{equation}
H_{ij} \equiv \frac{\partial^2 U(\B r_1,\B r_2\dots,\B r_N)}{\partial \B r_i \partial \B r_j} \ . \label{defH}
\end{equation}
Being a symmetric real matrix the eigenvalues $\{\lambda_i\}_{i=1}^{dN}$ of $\B H$ are real and
semi-positive as long as the system is mechanically stable. The normal modes of the system have frequencies
$\omega_i\equiv \sqrt{\lambda_i}$ from which the density of states can be calculated. Mechanical instabilities are associated
with non-zero eigenvalues going to zero, and the associated eigenfunction predicts the non-affine material response
during the instability \cite{06ML,10KLLP}.

Once the system gains thermal energy at temperature $T$, it is never in mechanical equilibrium since the
particles constantly dance around a mean position. Thus a temporal Hessian calculated at any given time will exhibit
negative eigenvalues and will be less useful in predicting instabilities or providing structural information \cite{18PNMESZ}.
Nevertheless, at temperatures low enough compared to the glass transition temperature $T_g$, the particles
are stuck for sufficiently long times in cages, allowing us to compute their average positions. It is therefore
interesting and relevant to attempt to characterize the average configurations and their instabilities
using a modified Hessian $\hat{\B H}$ which is adapted for the thermal conditions. In Sect.~\ref{candi} we discuss a number of candidates,
requiring a minimal condition that as long as the average positions are stationary and stable,
the eigenvalues of $\hat{\B H}$ were semipositive, in contradistinction with the instantaneous Hessian Eq.~(\ref{defH}). Having a number
of candidate, we test their efficacy in a standard model of a glass former which is described in Sect.~\ref{model}. In Sect.~\ref{Hessian}
we motivate the selection of one of the candidate Hessian which appears to provide an excellent characterization of the
plastic events in the average configurations.  In fact the non-affine displacement of the {\em average} particle positions
during a plastic event can be understood from the knowledge of the eigenvalues and eigenfunctions of $\hat{\B H}$
just before the event takes place.

Possibly one of the interesting conclusions of this study is that providing a typical glass forming
system with temperature is not like increasing the level of allowed motion on top of the athermal
energy landscape. The dynamics which temperature allows results in momentum transfers that are interpreted as
effective forces between the particles that are very different from the bare forces \cite{06BW,16GLPPRR,18PPPRS,19PPSZ}. Thus even
when the bare forces are binary, the dressed forces gain ternary, quaternary and higher order 
components due to multiple interactions between particles. If one thinks of these forces
as derivable from an effective Hessian as discuss below, it becomes obvious that the resulting 
``energy landscape" becomes quite different from the athermal counterpart, and explicitly temperature dependent. 
This important point is emphasized below in the conclusion section.

\section{Candidate Hessians}
\label{candi}

The first derivation of a microscopic theory of  the response of the materials to external mechanical strains was developed by Born (see, e.g., \cite{Born98,63MMWW}) for athermal systems. Further developments of the microscopic theory of elasticity at {\em finite temperatures}
using statistical mechanics were offered in Ref.~\cite{69SHH}. The result is that thermal corrections to the Born theory
include fluctuation terms, the magnitude of which are significant at high temperatures, vanishing at zero temperature in the case of perfect
crystals. A generalization of this approach to arbitrary systems in the solid state was developed in \cite{89Lut}. The limit of zero
temperatures in this approach indicates the existence of non-vanishing  fluctuation contributions in systems that are more complex than the perfect
crystal. Considering systems at zero temperature has an obvious advantage: it is possible to study the response of a single configuration
which is an ``inherent state" \cite{84SW}. This allows us to use a purely mechanical approach \cite{06ML,06MLa,10KLLP,11HKLP,12DKP,12DHP}
which is very useful in studying mechanical properties of amorphous solids.
The Hessian matrix, whose eigenvalues are semi-positive at $T=0$, provides important information, leading to an athermal theory that provides
good understanding of the density of states, of plastic events, and of the failure mechanisms of amorphous solids.

In order to lift the methods that
were so useful at $T=0$ to finite temperatures one needs to recognize that although the particle positions in a thermal systems are indeed not stationary,
in glasses with large relaxation times one can determine the averaged positions much before the onset of diffusion or the glass relaxation to
thermodynamic equilibrium. The averaged positions are obviously stationary, and can be used to determine the renormalized force laws that hold
these average positions stable \cite{06BW,16GLPPRR,18PPPRS,19PPSZ}. It was recognized that the renormalized forces are very different from the bare forces.
For example even if the bare forces are binary, the renormalized forces are not; generically they contain ternary, quaternary and higher order
contribution \cite{16GLPPRR,18PPPRS,19PPSZ}. The effective potential from which such renormalized forces can be derived is in general not known. We thus need to proceed with care in searching the ``correct" effective Hessian that may provide useful predictions for the plastic instabilities of the average configuration of a thermal glass
under external strains and stresses. In this section we demonstrate possible definitions of an effective Hessian at finite temperatures with the aim of
rationalizing instabilities in the average configurations.

Consider a glassy system composed of $N$ particles with time dependent positions $\{\B r_i(t)\}_{i=1}^N$ which is
endowed with a Hamiltonian $U\left(\B r_1(t), \cdots, \B r_N(t)\right)$. Assume that the system is in temperature $T$ which is sufficiently low
so that the glassy relaxation time (or the effective
diffusion time) $\tau_G$ is long enough, so that one can compute the time averaged positions $\B R_i$:
\begin{equation}
\B R_i \equiv \frac{1}{\tau}\int_0^\tau dt ~\B r_i(t) \ , \label{defRi}	
\end{equation}
where $\tau\ll \tau_G$.
By definition the positions $\B R_i$ are time independent and the configuration $\{\B R_i\}_{i=1}^N$ is stable, at least
for the time interval $[0, \tau_G]$. In addition to the mean positions we will employ below also the covariance matrix $\B \Sigma$ defined
as
\begin{equation}
\B \Sigma_{ij}\equiv \frac{1}{\tau}\int_0^\tau dt \left(\B r_i(t) - \B R_i\right)\left(\B r_j(t) - \B R_j\right) \ .
\end{equation}
We can then define three different effective Hessians as follows:

\subsection{Hessian computed at the average positions.}
	
To emphasize the fact that the bare Hessian which is computed from the bare potential (which is fully sufficient at
zero temperature) is {\em not} providing useful information at finite temperatures, we consider the first possible candidate Hessian, denoted as $\B H^{(1)}$.  Here 
derivatives of the bare potential are computed at the {\em average} positions:
\begin{equation}
\B H^{(1)}_{ij}= \frac{\partial U(\B r_1 \cdots \B r_N)} {\partial \B r_i\partial \B r_j}\Big |_{\B r_i=\B R_i,\B r_j =\B R_j} \ .
\end{equation}

\subsection{Hessian computed as the time average of the instantaneous Hessian}

Following the success of computing the effective forces in thermal glasses as the time average of the
instantaneous forces \cite{16GLPPRR,18PPPRS,19PPSZ}	one can try also to time average the instantaneous Hessian, denoting
it $\B H^{(2)}$:
\begin{equation}
\B H_{ij}^{(2)} \equiv \frac{1}{\tau}\int_0^\tau dt \frac{\partial^2 U(\B r_1(t),\B r_2(t)\dots,\B r_N(t))}{\partial \B r_i(t) \partial \B r_j(t)} \ ,
\end{equation}
where the integration time $\tau$ is large enough to achieve convergence and small enough for the cage structure to remain
intact. One could hope that this candidate Hessian would be sensitive to the dynamics to be able to provide useful information. 

\subsection{Hessian computed from the inverse of the covariance matrix}

The third candidate Hessian is the most tricky, since it is based on the notion of effective potential. As said above,
in amorphous solids we can define the average positions $\B R_i$ which are stationary in time, and can consider
the effective forces $\hat {\B  F}$  that stabilize these positions. These forces are determined by the momentum transfer during
the dynamics, but the momentum transferred between particles $i$ and $j$ can depend on intervening interaction between
particles $i$ and $k$ (leading to ternary rather than binary effective interactions), or between particle $i$, $k$
and $\ell$, giving rise to quaternary effective interactions etc. While the {\em bare} force on the $i$th particle $-\partial U/\partial \B r_i$
does not vanish at $\B R_i$, we can think of an effective Hamiltonian $\hat U$ from which (in principle) the effective forces $\hat {\B  F}_i$
can be derived, and of course should satisfy the condition $-\partial\hat  U/\partial \B r_i=0$ when computed at $\B R_i$.

Although we do not have an explicit solution for the effective potential $\hat U$, we can still Taylor expand it around the
average positions $\B R_i$:
\begin{equation}
\hat U(\{\B r_i\}) =\hat U(\{\bf R_i\})+\frac{1}{2}\tilde{{\bf u}}{\bf H}^{(3)}\large|_{\bf R}{\bf u}+\cdots,
\label{TE}
\end{equation}
where ${\bf u}=\{\B r_i-\B R_i\}$ is the set of particle displacements from the average. The candidate Hessian
$\B H^{(3)}$ is given by
\begin{equation}
\B H_{ij}^{(3)}=\frac{\partial^2 \hat U(\{\bf r_i\})}{\partial \B r_{i}\partial \B r_{j}}\large|_{\bf R_i,\B R_j}\  .
\label{Hess3}
\end{equation}

Of course, at this point the definition is useless since we do not know the actual form of $\hat U({\bf r_i})$. But if we did
we could also compute the average positions and the covariance matrix according to
\begin{equation}
\B R_i=\frac{\int \B r_i e^{-\frac{\hat U(\{\bf r_j\})}{k_BT}} d\{\bf r_j\}}{\int e^{-\frac{\hat U(\{\B  r_j\})}{k_BT}} d\{\bf r_j\}}
\label{MCtrack}
\end{equation}
\begin{equation}
\langle \B r_i\B r_j\rangle=\frac{\int \B r_i \B r_j e^{-\frac{\hat U(\{\bf r_k\})}{k_BT}} d\{\bf  r_k\}}{\int e^{-\frac{\hat U(\{\B r_k\})}{k_BT}} d\{\bf r_k\}}
\label{SecMom}
\end{equation}
Knowing the first and the second moments one can define the multivariate Gaussian distribution
\begin{equation}
f({\B r})=C \exp\bigg\{-\frac{1}{2}\widetilde{(\B  r-\B  R)}{\B \Sigma}^{-1}(\B  r-\B  R)\bigg\} \ .
\label{mv1}
\end{equation}
Comparing Eq.~(\ref{TE}) and Eq.~(\ref{mv1}) one identifies
\begin{equation}
{\bf H}^{(3)}=k_B T {\bf \Sigma}^{-1}.
\label{PDh}
\end{equation}
We should note at this point that the covariance matrix is positive semidefinite only, due to the existence of
Goldstone modes,
and Eq.~(\ref{PDh}) should be replaced by
\begin{equation}
{\bf H}^{(3)}=k_B T {\bf \Sigma}^+,
\label{PDhPI}
\end{equation}
where ${\bf \Sigma}^+$ is the pseudo inverse of the covariance matrix.
Of course, the effective Hessian given by Eq.~(\ref{PDhPI})  and the covariance matrix have the same set of eigenfunctions
\begin{equation}
{\bf H}^{(3)}{\bf \Psi}_i=\lambda^H_i {\bf \Psi}_i
\label{effeig}
\end{equation}
and their eigenvalues are related by
\begin{equation}
\lambda^H_i=\frac{k_B T}{\lambda_i^{\Sigma^+}} \ .
\label{EigV}
\end{equation}
This approach was used to study the vibrational spectra of colloids and granular systems based on measurements using video and  confocal microscopy (see e.g., \cite{10GCDKB,12HBD}),
and restoration of effective interaction potential using simulation methods \cite{16SM}. It should be stressed that this method appears to have been used only
for unstrained systems. Below we will use this effective Hessian for shear strained systems before and after plastic responses. 

In the next section we will use a standard model of glass formers to decide which of the candidate Hessians provides
the information that we seek.

\section{Model}
\label{model}

To make a choice between the three candidate Hessians we compute them together with their eigenvalues and
eigenfunctions in a standard  model of a glass former, i.e. a binary
 mixture of point particles interacting via inverse power-law potentials \cite{99PH}. 50\% of the particles
 are ``small" (type $A$) and the other 50\% of the particles are ``large" (type $B$). The interaction
 between particle $\alpha$ (being $A$ or $B$) and particle $\beta$ (being $A$ or $B$)
are defined as
\begin{equation}
  \phi_{\alpha\beta}(r)= \epsilon \Big(\frac{\sigma_{\alpha\beta}}{r}\Big)^{12}\ .
\end{equation}
It is convenient to introduce reduced units, with $\sigma_{AA}=1$  being the units of length and
$\epsilon=1$ the unit of energy (with Boltzmann's constant being unity).
 \begin{figure}
\includegraphics[width=.45\textwidth]{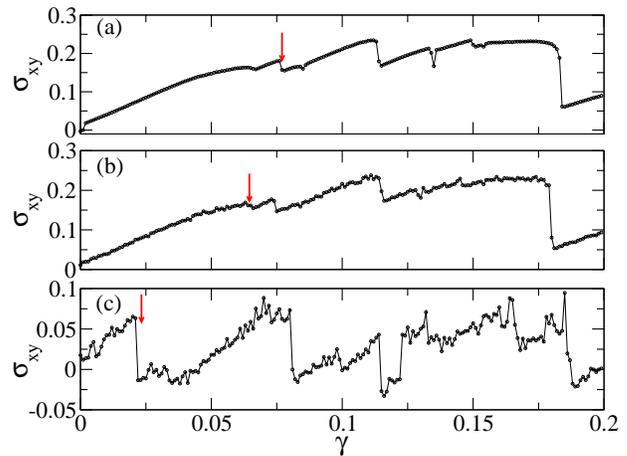}
\caption{The stress averaged over 10,000 Monte Carlo sweeps vs. strain, for three different temperatures; panel a: $T=0.0001$; panel b: $T=0.01$;
panel c: $T=0.1$. Arrows indicate the plastic events in the {\em average} configuration that are analyzed in detail below.}
\label{stressvsstrain}
 \end{figure}
 \begin{figure}
 	\vskip 0.5 cm
\includegraphics[width=.40\textwidth]{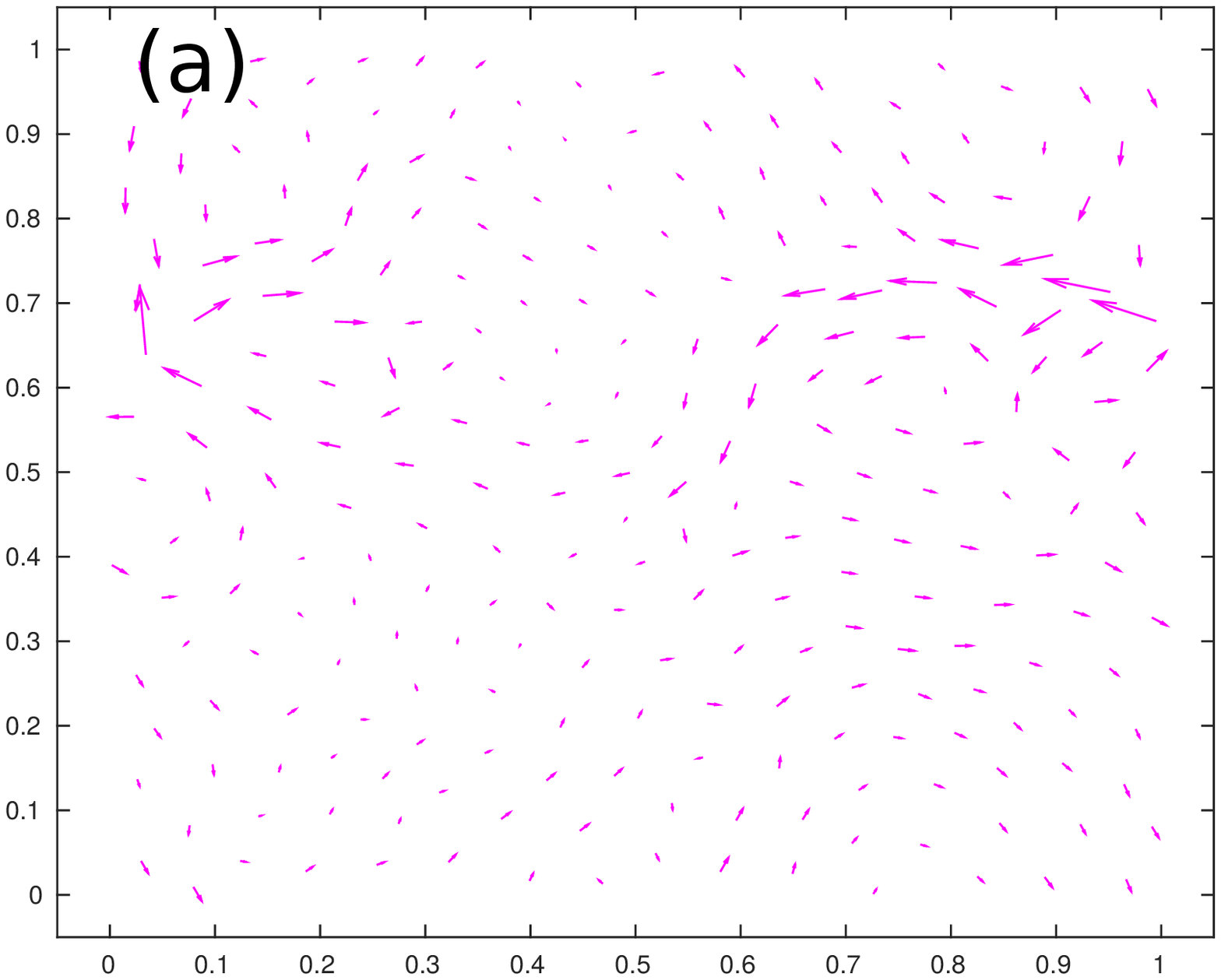}
\includegraphics[width=.40\textwidth]{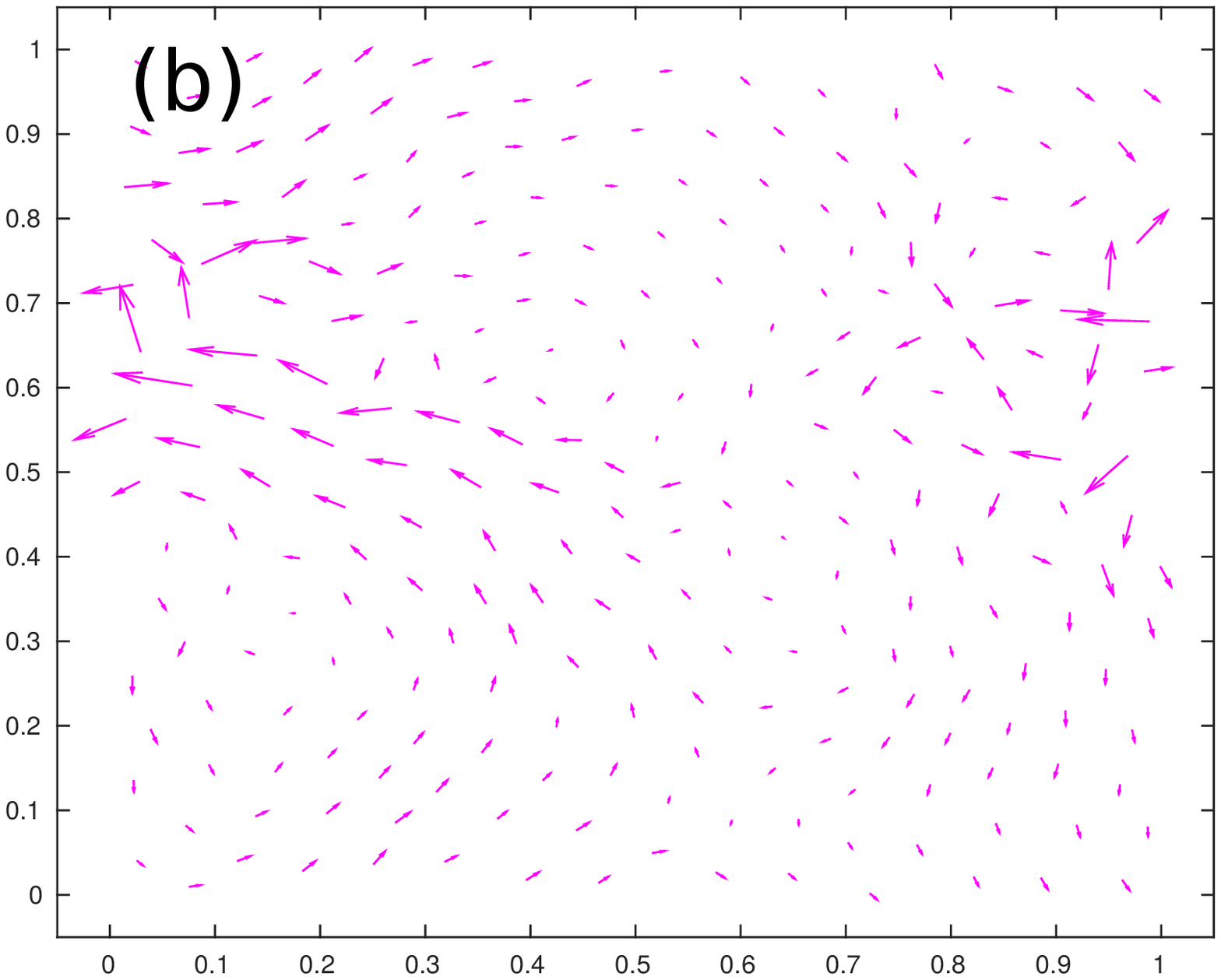}
\includegraphics[width=.40\textwidth]{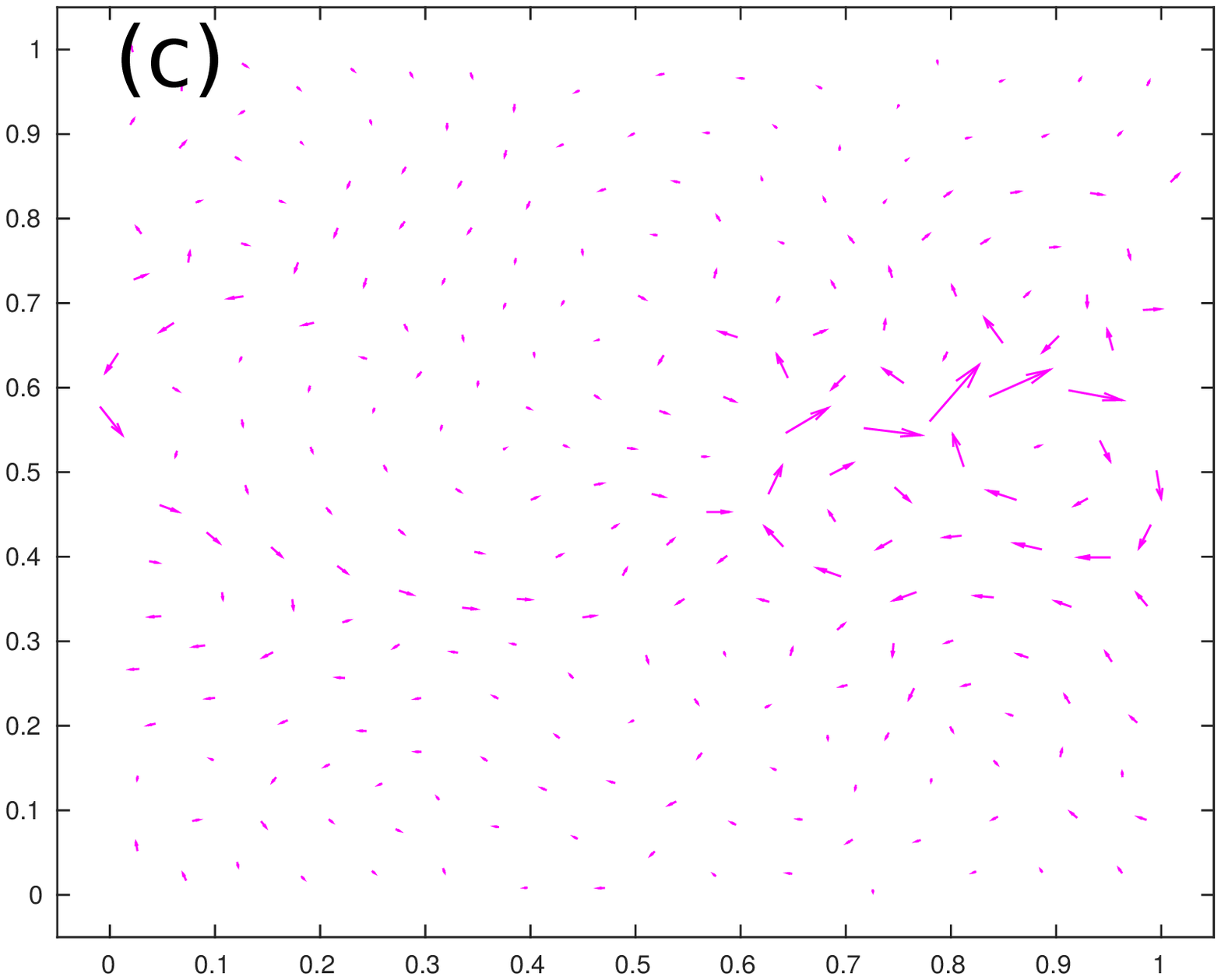}
 \caption{The non-affine displacement field for $T=0.0001$ (panel a), $T=0.01$ (panel b) and $T=0.1$ (panel c) measured for the time-averaged mean particle positions during the plastic drops indicated with arrows in Fig.~\ref{stressvsstrain}.  }
 \label{nonaffine}
 \end{figure}

\subsection{Time-averaged configurations and their instabilities}

Simulations were performed using a Monte Carlo method
in an NVT ensemble of $N=256$ particle  in a two dimensional box
of size $L\times L$ with periodic boundary conditions. $L$ was chosen such that at any temperature the density $\rho=0.76$ \cite{99PH}. The acceptance rate was
chosen to be $30 \%$ at all temperatures. We first equilibrate a system  at a temperature $T = 3$ and
then cool it down in steps of $\Delta T=10^{-3}$ to a target temperature $0<T\le 0.1$, where the upper limit
was chosen since in this system $T_g\approx 0.3$ \cite{99PH}. Once the system is equilibrated at the target temperature
we begin to strain the system by simple shear in a quasi-static manner.  Here ``quasi-static" means that after every small step of strain
we allow the system to equilibrate performing 10,000 Monte Carlo sweeps. After completing the last Monte-Carlo sweep a small affine increase in  strain $\Delta\gamma$ is defined by
the volume preserving transformation
\begin{eqnarray}
  x^{\prime}_i&=&x_i+\Delta\gamma  y_i\nonumber \\
  y^{\prime}_i&=&y.
  \label{simpleshear}
  \end{eqnarray}
In thermal glasses every such affine step destroys the thermal equilibrium, causing a change in the average positions and the covariance matrix.
To regain equilibrium one should allow a nonaffine relaxation to a new average particle positions and covariance matrix.  In the quasi-static protocol
we make sure that the average positions are stabilized. Computing the average stress is done in a subsequent run of $1,500,000$ sweeps, again
controlling convergence. Typical time-averaged stress $\sigma_{xy}$ vs strain $\gamma$ for temperatures $T=0.001$, $T=0.01$ and $T=0.1$
 are shown in Fig.~\ref{stressvsstrain}.

 We note that the time-averaged stress exhibits sharp drops even at $T/T_g\approx 0.33$, and these
 are easily distinguishable from temperature fluctuations that are averaged out in these plots. We refer to the
 drops in the time-averaged stress as plastic drops. Our aim
 here is to understand, and possibly predict,  the non-affine change in the {\em average} positions of particles (average displacement) that occur during these plastic drops,
 using the Hessian and its eigen-properties {\em before the drop takes place}.

 To prepare for comparisons with theory we consider the non-affine displacement fields that are obtained from the time averaged configurations before and after a sharp drop. These displacement fields
 are obtained by subtracting from the measured change in averaged configurations the last affine step
 Eq.~(\ref{simpleshear}) before the drop. We denote below the {\em normalized} non-affine displacement field as ${\bf {\cal N}}(\B r)$. Examples of such fields are shown in Fig.~\ref{nonaffine}
 for $T=0.0001$ (panel a), $T=0.01$ (panel b) and $T=0.1$ (panel c). The displacement fields shown here are associated with the plastic drops that are indicated with an arrow in Fig.~\ref{stressvsstrain}.

We note that the observable sharp plastic drops in the time-averaged stress can be accompanied by non-affine displacement in
the average configurations that can be either
system spanning or localized.  In our simulation systems spanning events are more prevalent, presumably smaller localized drops are swamped by temperature fluctuations.

\section{Which Hessian?}
\label{Hessian}

At this point we are ready to select which Hessian fits the bill and provides a useful theory for understanding
the instabilities and the non-affine displacements in the average configurations. For the present model, and
presumably quite generally, the first candidate $\B H^{(1)}$ can be ruled out quite immediately since its calculation
at the average points yields negative eigenvalues similarly to the instantaneous Hessian. We thus do not discuss it further.

The second candidate $\B H^{(2)}$ computed as the time average of the bare Hessian provides a semi-positive spectrum
at all the simulated temperatures, with 2 zero eigenvalues for the Goldstone modes. It therefore appears on the face
of it to be a valid candidate. However, computing the eigenvalues of $\B H^{(2)}$ at a series of increasing temperatures
reveals that even in the equilibrium state with zero strain these eigenvalues {\em increase} as a function of temperature.
We propose that this is highly unphysical. One expects the amorphous solid to become softer as temperature increases,
with eigenvalues going to zero at the glass transition temperature. We have checked that this unphysical behavior persists
for different model Hamiltonians, including the standard Lennard-Jones glass \cite{94KA}. A second problem with $\B H^{(2)}$ is that
its lowest eigenvalue never appears to approach zero near a plastic instability. Its eigenfunctions did not show resemblance
to the non-affine events that we care to describe.

In hindsight, this should not be surprising. The bare Hessian has information about the binary interactions only, and by time averaging it we do not input anywhere the pertinent information about higher order interactions. Not having this information makes the second candidate useless.  We thus discard this candidate Hessian from our list.

The third candidate Hessian $\B H^{(3)}$ appears to provide us with the best guide for understanding the plastic instabilities
as we demonstrate in detail in the next section. But it also agrees with physical intuition, providing  eigenvalues
that {\em reduce} with increasing temperature, exhibiting a semi-positive spectrum with two Goldstone modes. We thus drop from this
point onward the superscript (3) and denote the chosen candidate Hessian as $\hat {\B H}$ to distinguish it from the bare Hessian $\B H$.

 \begin{figure}
\includegraphics[width=.45\textwidth]{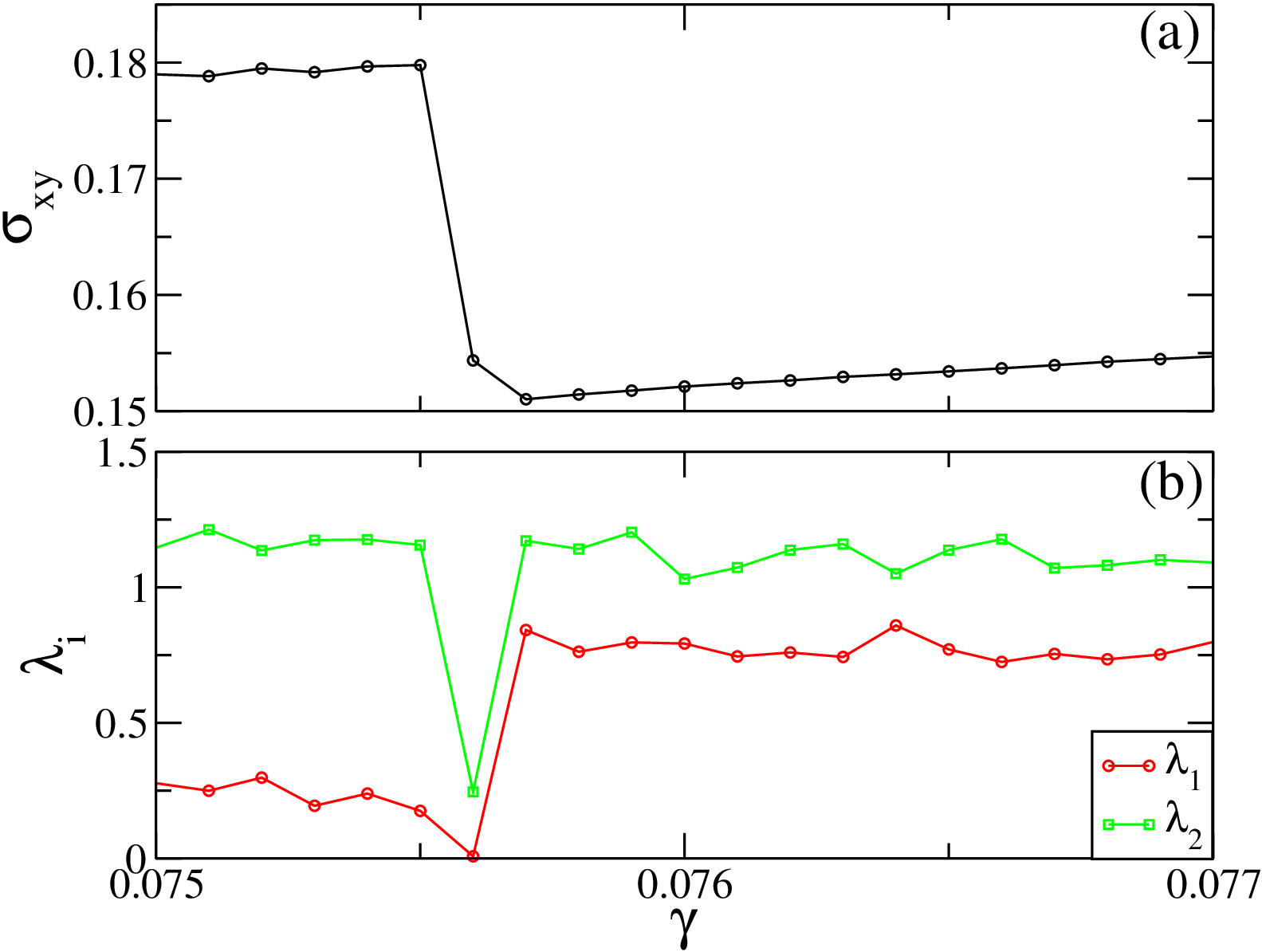}
 \caption{Panel a: the drop in the average stress expanded at $\gamma\approx 0.0755$. Panel b: the 2 lowest lying eigenvalues of the average Hessian $\hat {\B H}$ as a function of strain in the expanded strain neighborhood of the sharp drop indicated with an arrow in Fig.~\ref{stressvsstrain}. The temperature is $T=0.0001$.}
 \label{eigstrain1}
 \end{figure}
 \begin{figure}
 	\includegraphics[width=.45\textwidth]{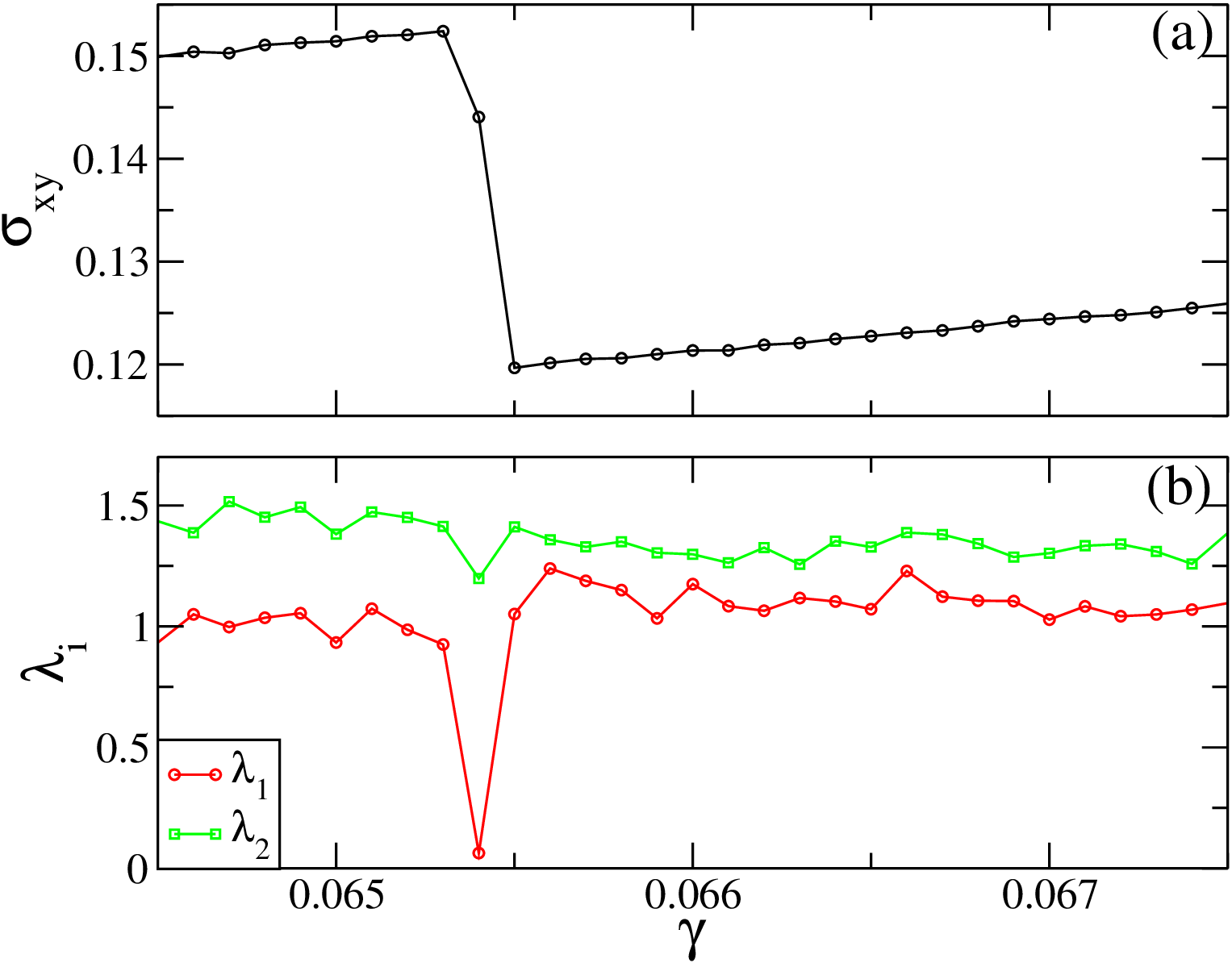}
 	\caption{Panel a: the drop in the average stress expanded at $\gamma\approx 0.0653$. Panel b: the 2 lowest lying eigenvalues of the average Hessian $\hat {\B H}$ as a function of strain in the expanded strain neighborhood of the sharp drop indicated with an arrow in Fig.~\ref{stressvsstrain}. The temperature is $T=0.01$.}
 	\label{eigstrain2}
 \end{figure}
 \begin{figure}
 	\includegraphics[width=.45\textwidth]{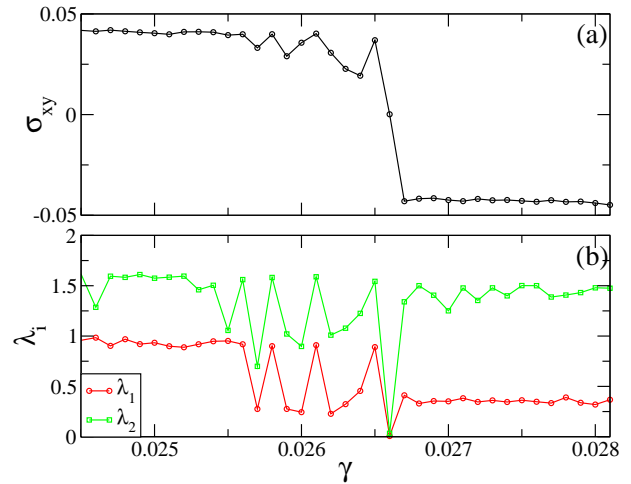}
 	\caption{Panel a: the drop in the average stress expanded at $\gamma\approx 0.0265$. Panel b: the 2 lowest lying eigenvalues of the average Hessian $\hat {\B H}$ as a function of strain in the expanded strain neighborhood of the sharp drop indicated with an arrow in Fig.~\ref{stressvsstrain}. The temperature is $T=0.1$.}
 	\label{eigstrain3}
 \end{figure}
 \begin{figure}[h!]
 	\includegraphics[width=.40\textwidth]{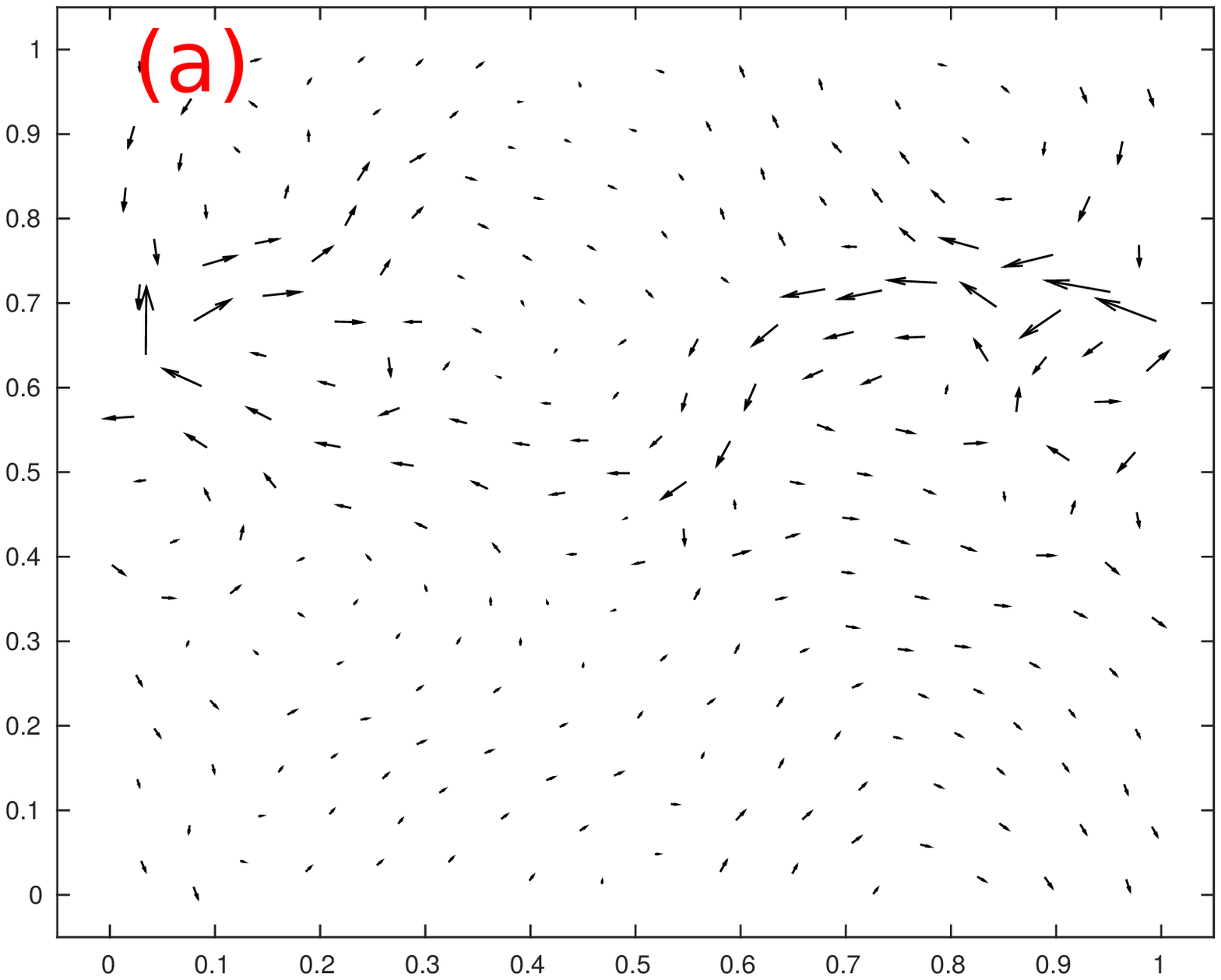}
 	\includegraphics[width=.40\textwidth]{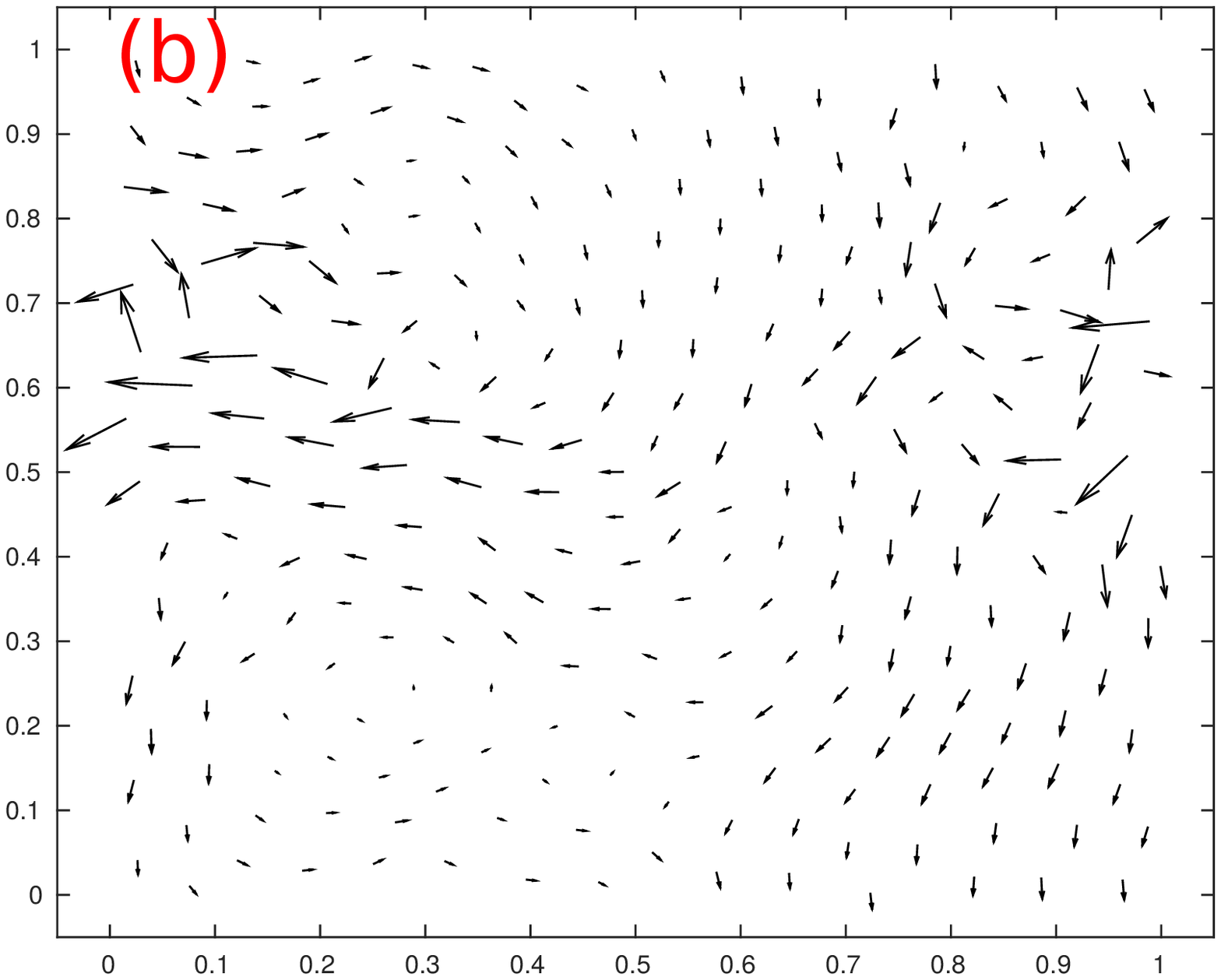}
 	\includegraphics[width=.40\textwidth]{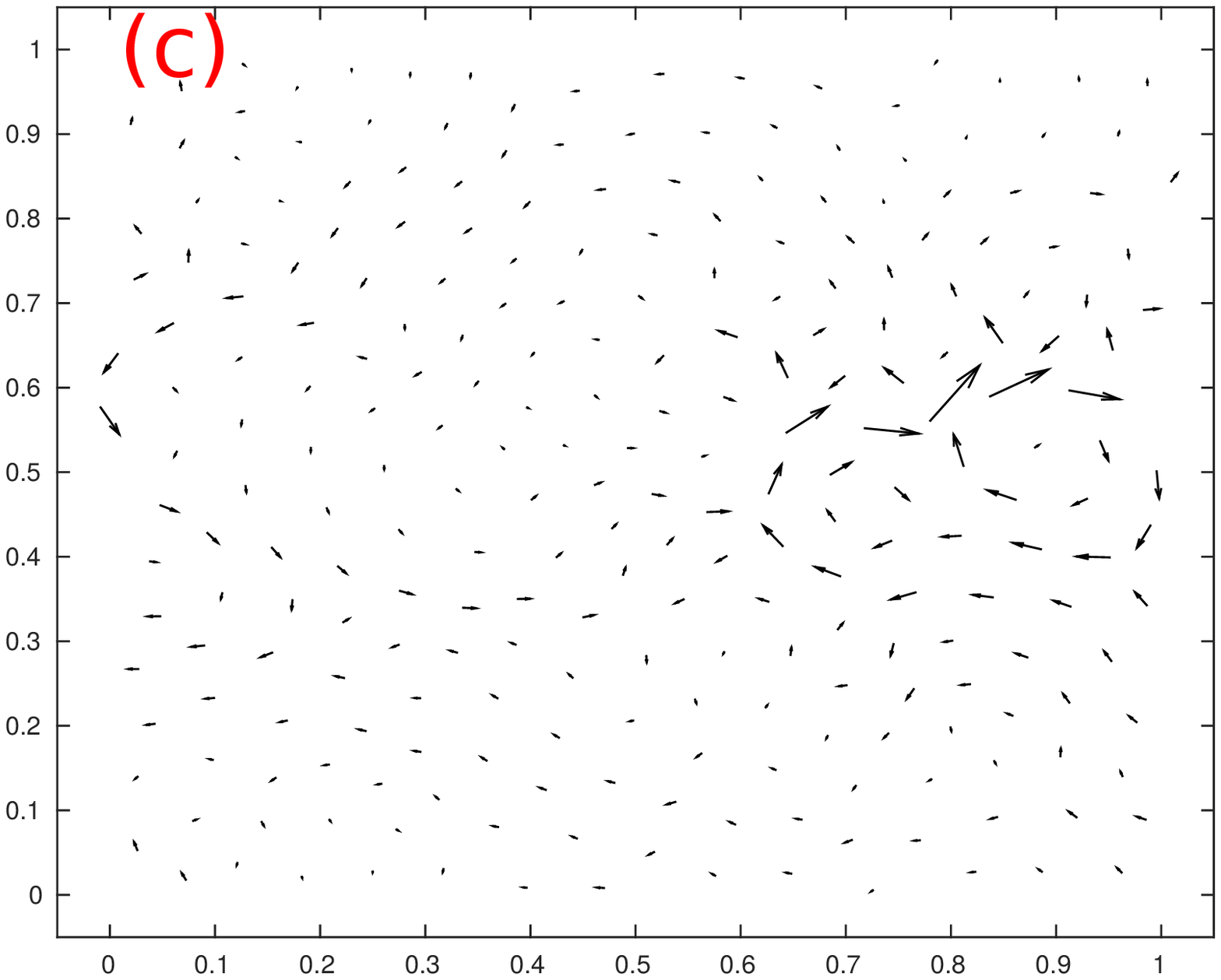}
 	\caption{The eigenfunctions of $\hat {\B H}$ associated with the lowest eigenvectors that tend to vanish
 		at the plastic instability. Their visual resemblance to the actual non-affine displacement fields shown
 		in Figs.~\ref{nonaffine} is clear, and is emphasized further in Fig.~\ref{comparison}.}
 	\label{eigenfunctions}
 \end{figure}

\section{Predicting non-affine time-averaged displacements}

The most stringent test on the chosen Hessian is its ability to predict the nonaffine-displacement using its eigenfunctions and eigenvalues {\em before} the instability takes place. As said above, excluding Goldstone modes,
the eigenvalues of $\hat {\B H}$ are all real and positive as long as the system is stable.
Moreover, they display a very weak dependence on the strain as long as the instability is
not approached. An example of the strain-dependence stress and of the 2 lowest lying eigenvalues of $\hat {\B H}$
are shown in Fig.~\ref{eigstrain1}-\ref{eigstrain3} for all the three temperatures in Fig.~\ref{stressvsstrain}.
We see that there is a clear tendency for the lowest positive eigenvalue to vanish at strain values that correspond to the sharp stress drops.

\begin{figure}[h!]
	\includegraphics[width=.40\textwidth]{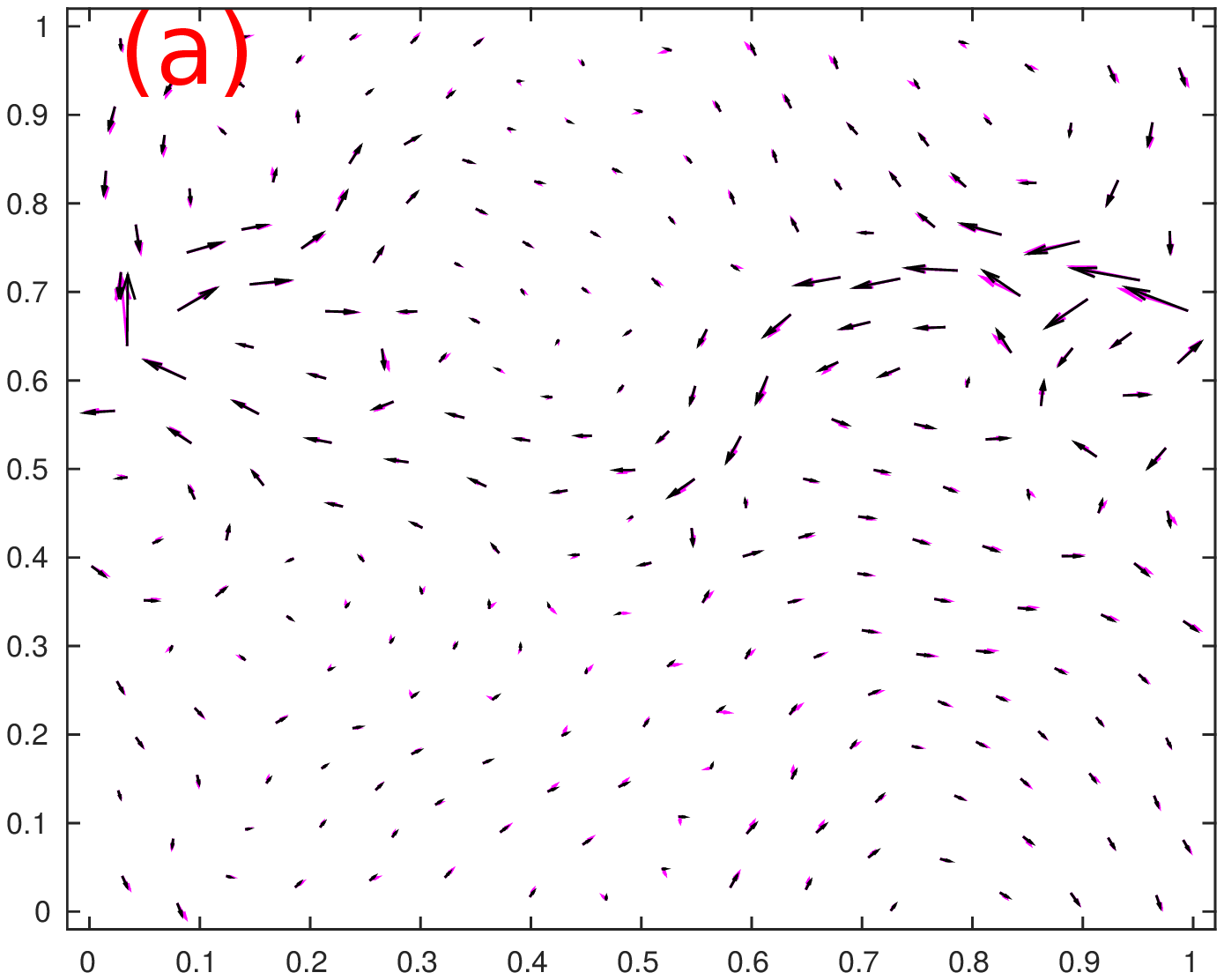}
	\includegraphics[width=.40\textwidth]{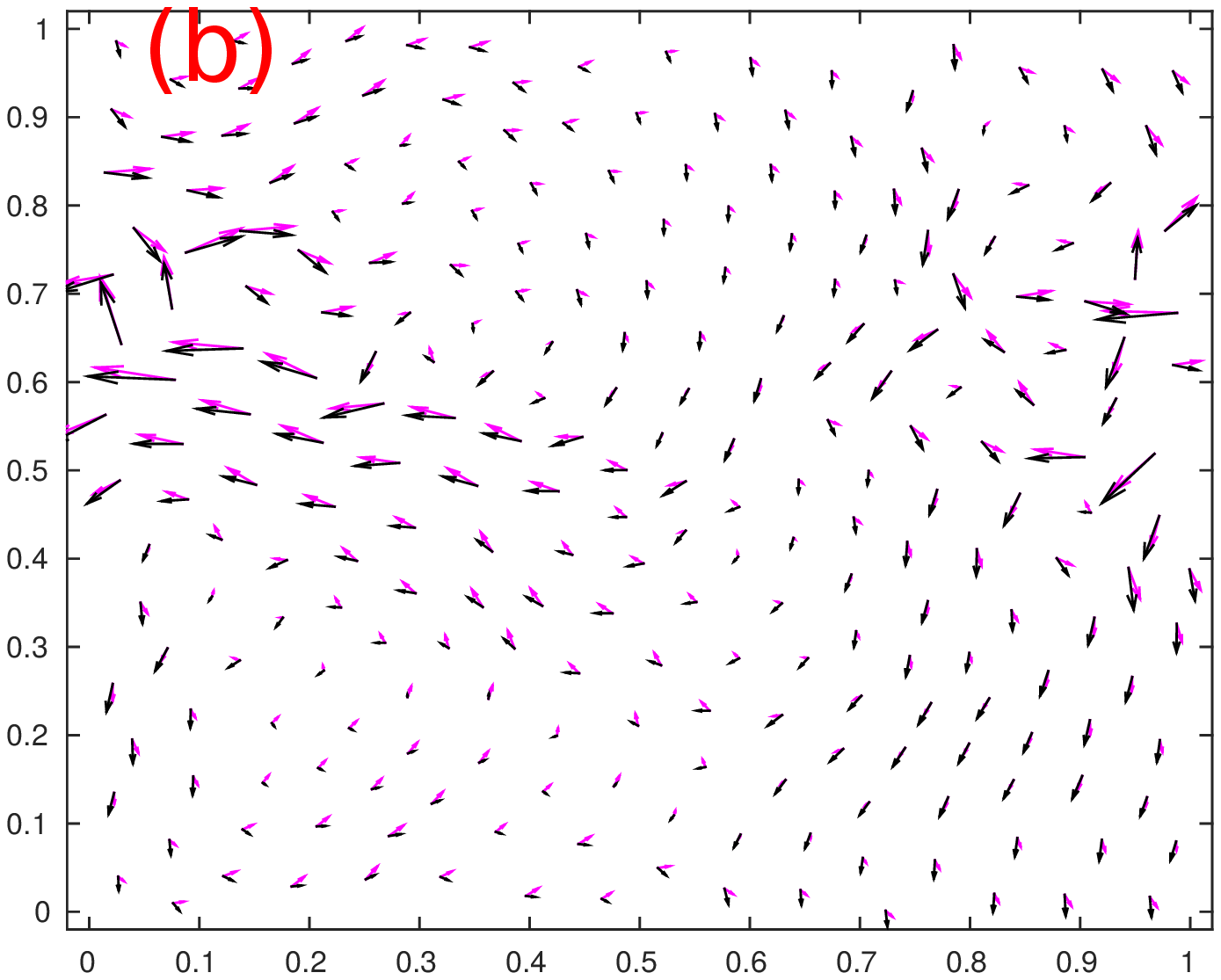}
	\includegraphics[width=.40\textwidth]{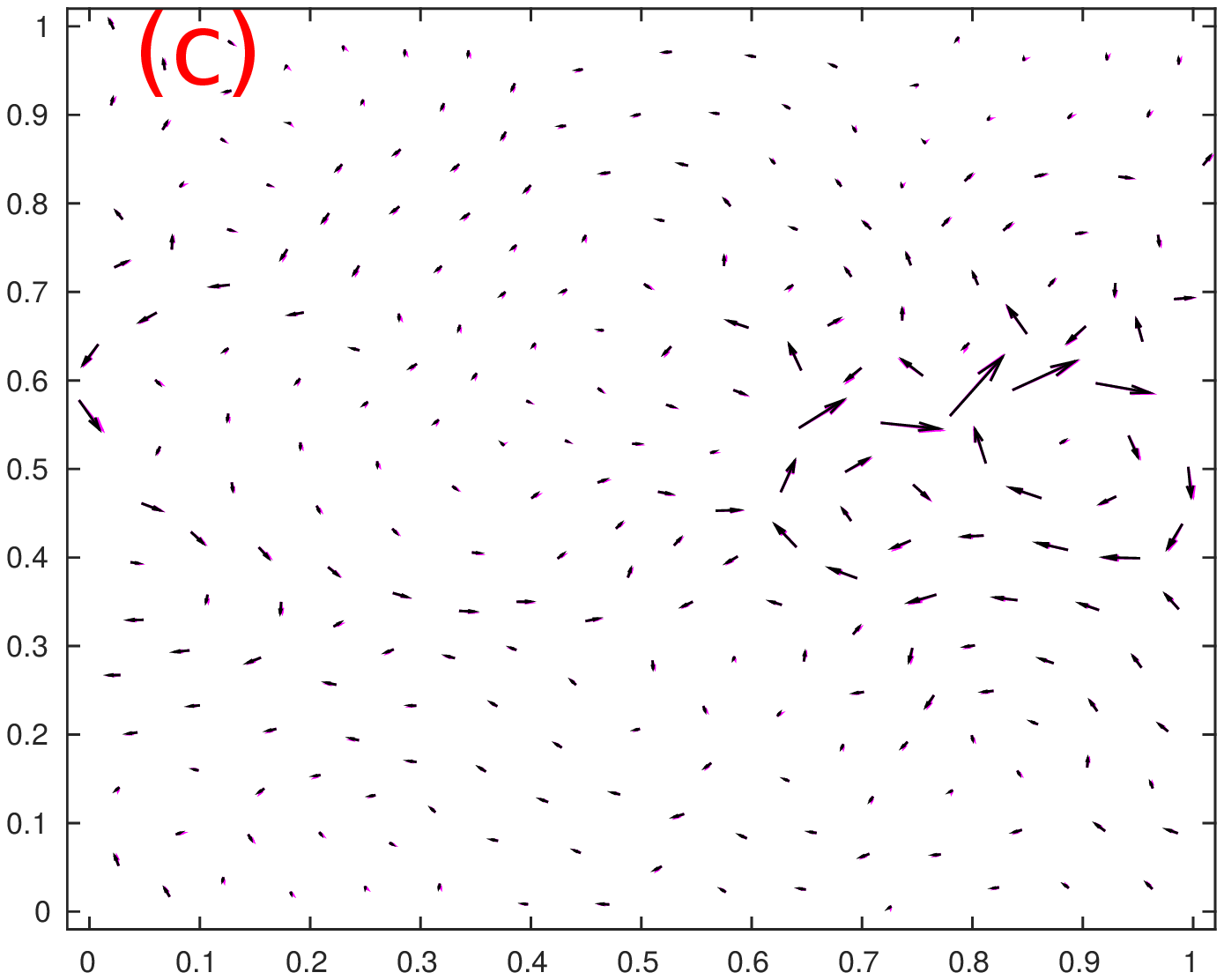}
	\caption{A comparison between the eigenfunctions and the actual non-affine displacement fields of the average configurations. Here we superimposed 
		the non-affine displacement fields shown
		in Figs.~\ref{nonaffine} with the eigenfunctions shown in Fig.~\ref{eigenfunctions}. Their visual resemblance  is striking.}
	\label{comparison}
\end{figure}

At this point consider the eigenfunctions $\B \Psi_1$ of $\hat {\B H}$  which are associated with the lowest-lying
eigenvalues $\lambda_1$ at the three temperatures discussed here. In Fig.~\ref{eigenfunctions} we show the three eigenfunctions associated with the lowest eigenvalue at each temperature, and these should be compared with the non-affine displacement fields
in Fig.~\ref{nonaffine}. It is obvious to the eye that the correspondence is quite striking. To emphasize this correspondence we shown in Fig.~\ref{comparison} the non-affine
displacements superimposed on the eigenfunctions. A quantitative measure of the agreement (and the predictability of the non-affine displacement
field form the eigenfunction) is obtained by normalizing the displacement field and computing the
scalar product $a_1\equiv {\bf {\cal N}}(\B r)\cdot \B \Psi_1$ with the (orthonormal) eigenfunction. For the three temperatures discussed here
we find the scalar product to be 0.981 for $T=0.0001$,  0.893 for  $T=0.01$ and 0.992 for   $T=0.1$.  Obviously, the eigenfunction associated
with the eigenvalue that goes to zero at the instability are able to predict the non-affine displacement field
in the {\em average} position quite successfully. It is very interesting that this predictive capability exists even
at $T=0.1$ which is about a third of $T_g$.
\section{Summary and Conclusions}

In this paper we examined the lifting of athermal Hessian methods to glasses at finite temperatures. The main idea is that at temperatures that are low enough one can determine a stationary average positions of all the involved particles.
Such a configuration has a vanishing effective force on each particle, exactly as at $T=0$ in an inherent state.
A well chosen Hessian matrix then should supply information about plastic instabilities that is as useful as the bare
Hessian at $T=0$. A-priori it is not obvious how to define such a ``well chosen" Hessian. We have examined in the
present paper a number of candidates, and discovered that the effective Hessian that is found from inverting
the covariance matrix is the most appropriate. It exhibits eigenvalue that decrease upon increasing temperatures,
in agreement with the expectation that the glassy solid softens up upon heating. More importantly, the lowest
eigenvalues of this Hessian tend to vanish at the strain values where the {\em average} stress suffers sharp
drops indicating an important plastic rearrangement in the {\em average} positions of the particles.
We discovered that the eigenfunctions associated with the lowest eigenvalue have almost full projection
on the nonaffine displacement field, providing us with a useful predictability of the non-affine plastic
event.

Having in mind the usefulness of the Hessian methods at athermal conditions, we trust that the present
results should have further implication on the study and understanding of the mechanical properties of thermal glasses.

\acknowledgments
This work had been supported in part by the the Joint Laboratory on ``Advanced and Innovative Materials" - Universita' di Roma ``La Sapienza" - WIS and by the US-Israel Binational Science Foundation. The authors are grateful
to Smarajit Karmakar and Vishnu Krishnan for useful discussions at the beginning of this project.

\bibliography{ALL}

\begin{thebibliography}{21}%
\makeatletter
\providecommand \@ifxundefined [1]{%
 \@ifx{#1\undefined}
}%
\providecommand \@ifnum [1]{%
 \ifnum #1\expandafter \@firstoftwo
 \else \expandafter \@secondoftwo
 \fi
}%
\providecommand \@ifx [1]{%
 \ifx #1\expandafter \@firstoftwo
 \else \expandafter \@secondoftwo
 \fi
}%
\providecommand \natexlab [1]{#1}%
\providecommand \enquote  [1]{``#1''}%
\providecommand \bibnamefont  [1]{#1}%
\providecommand \bibfnamefont [1]{#1}%
\providecommand \citenamefont [1]{#1}%
\providecommand \href@noop [0]{\@secondoftwo}%
\providecommand \href [0]{\begingroup \@sanitize@url \@href}%
\providecommand \@href[1]{\@@startlink{#1}\@@href}%
\providecommand \@@href[1]{\endgroup#1\@@endlink}%
\providecommand \@sanitize@url [0]{\catcode `\\12\catcode `\$12\catcode
  `\&12\catcode `\#12\catcode `\^12\catcode `\_12\catcode `\%12\relax}%
\providecommand \@@startlink[1]{}%
\providecommand \@@endlink[0]{}%
\providecommand \url  [0]{\begingroup\@sanitize@url \@url }%
\providecommand \@url [1]{\endgroup\@href {#1}{\urlprefix }}%
\providecommand \urlprefix  [0]{URL }%
\providecommand \Eprint [0]{\href }%
\providecommand \doibase [0]{http://dx.doi.org/}%
\providecommand \selectlanguage [0]{\@gobble}%
\providecommand \bibinfo  [0]{\@secondoftwo}%
\providecommand \bibfield  [0]{\@secondoftwo}%
\providecommand \translation [1]{[#1]}%
\providecommand \BibitemOpen [0]{}%
\providecommand \bibitemStop [0]{}%
\providecommand \bibitemNoStop [0]{.\EOS\space}%
\providecommand \EOS [0]{\spacefactor3000\relax}%
\providecommand \BibitemShut  [1]{\csname bibitem#1\endcsname}%
\let\auto@bib@innerbib\@empty
\bibitem [{\citenamefont {Lema{\^\i}tre}\ and\ \citenamefont
  {Maloney}(2006)}]{06ML}%
  \BibitemOpen
  \bibfield  {author} {\bibinfo {author} {\bibfnamefont {A.}~\bibnamefont
  {Lema{\^\i}tre}}\ and\ \bibinfo {author} {\bibfnamefont {C.}~\bibnamefont
  {Maloney}},\ }\href@noop {} {\bibfield  {journal} {\bibinfo  {journal}
  {Journal of statistical physics}\ }\textbf {\bibinfo {volume} {123}},\
  \bibinfo {pages} {415} (\bibinfo {year} {2006})}\BibitemShut {NoStop}%
\bibitem [{\citenamefont {Karmakar}\ \emph {et~al.}(2010)\citenamefont
  {Karmakar}, \citenamefont {Lema{\^{i}}tre}, \citenamefont {Lerner},\ and\
  \citenamefont {Procaccia}}]{10KLLP}%
  \BibitemOpen
  \bibfield  {author} {\bibinfo {author} {\bibfnamefont {S.}~\bibnamefont
  {Karmakar}}, \bibinfo {author} {\bibfnamefont {A.}~\bibnamefont
  {Lema{\^{i}}tre}}, \bibinfo {author} {\bibfnamefont {E.}~\bibnamefont
  {Lerner}}, \ and\ \bibinfo {author} {\bibfnamefont {I.}~\bibnamefont
  {Procaccia}},\ }\href {\doibase 10.1103/PhysRevLett.104.215502} {\bibfield
  {journal} {\bibinfo  {journal} {Phys. Rev. Lett.}\ }\textbf {\bibinfo
  {volume} {104}},\ \bibinfo {pages} {215502} (\bibinfo {year}
  {2010})}\BibitemShut {NoStop}%
\bibitem [{\citenamefont {Palyulin}\ \emph {et~al.}(2018)\citenamefont
  {Palyulin}, \citenamefont {Ness}, \citenamefont {Milkus}, \citenamefont
  {Elder}, \citenamefont {Sirk},\ and\ \citenamefont {Zaccone}}]{18PNMESZ}%
  \BibitemOpen
  \bibfield  {author} {\bibinfo {author} {\bibfnamefont {V.~V.}\ \bibnamefont
  {Palyulin}}, \bibinfo {author} {\bibfnamefont {C.}~\bibnamefont {Ness}},
  \bibinfo {author} {\bibfnamefont {R.}~\bibnamefont {Milkus}}, \bibinfo
  {author} {\bibfnamefont {R.~M.}\ \bibnamefont {Elder}}, \bibinfo {author}
  {\bibfnamefont {T.~W.}\ \bibnamefont {Sirk}}, \ and\ \bibinfo {author}
  {\bibfnamefont {A.}~\bibnamefont {Zaccone}},\ }\href {\doibase
  10.1039/C8SM01468J} {\bibfield  {journal} {\bibinfo  {journal} {Soft Matter}\
  }\textbf {\bibinfo {volume} {14}},\ \bibinfo {pages} {8475} (\bibinfo {year}
  {2018})}\BibitemShut {NoStop}%
\bibitem [{\citenamefont {Brito}\ and\ \citenamefont {Wyart}(2006)}]{06BW}%
  \BibitemOpen
  \bibfield  {author} {\bibinfo {author} {\bibfnamefont {C.}~\bibnamefont
  {Brito}}\ and\ \bibinfo {author} {\bibfnamefont {M.}~\bibnamefont {Wyart}},\
  }\href@noop {} {\bibfield  {journal} {\bibinfo  {journal} {EPL (Europhysics
  Letters)}\ }\textbf {\bibinfo {volume} {76}},\ \bibinfo {pages} {149}
  (\bibinfo {year} {2006})}\BibitemShut {NoStop}%
\bibitem [{\citenamefont {Gendelman}\ \emph {et~al.}(2016)\citenamefont
  {Gendelman}, \citenamefont {Lerner}, \citenamefont {Pollack}, \citenamefont
  {Procaccia}, \citenamefont {Rainone},\ and\ \citenamefont
  {Riechers}}]{16GLPPRR}%
  \BibitemOpen
  \bibfield  {author} {\bibinfo {author} {\bibfnamefont {O.}~\bibnamefont
  {Gendelman}}, \bibinfo {author} {\bibfnamefont {E.}~\bibnamefont {Lerner}},
  \bibinfo {author} {\bibfnamefont {Y.~G.}\ \bibnamefont {Pollack}}, \bibinfo
  {author} {\bibfnamefont {I.}~\bibnamefont {Procaccia}}, \bibinfo {author}
  {\bibfnamefont {C.}~\bibnamefont {Rainone}}, \ and\ \bibinfo {author}
  {\bibfnamefont {B.}~\bibnamefont {Riechers}},\ }\href {\doibase
  10.1103/PhysRevE.94.051001} {\bibfield  {journal} {\bibinfo  {journal} {Phys.
  Rev. E}\ }\textbf {\bibinfo {volume} {94}},\ \bibinfo {pages} {051001}
  (\bibinfo {year} {2016})}\BibitemShut {NoStop}%
\bibitem [{\citenamefont {Parisi}\ \emph {et~al.}(2018)\citenamefont {Parisi},
  \citenamefont {Pollack}, \citenamefont {Procaccia}, \citenamefont {Rainone},\
  and\ \citenamefont {Singh}}]{18PPPRS}%
  \BibitemOpen
  \bibfield  {author} {\bibinfo {author} {\bibfnamefont {G.}~\bibnamefont
  {Parisi}}, \bibinfo {author} {\bibfnamefont {Y.~G.}\ \bibnamefont {Pollack}},
  \bibinfo {author} {\bibfnamefont {I.}~\bibnamefont {Procaccia}}, \bibinfo
  {author} {\bibfnamefont {C.}~\bibnamefont {Rainone}}, \ and\ \bibinfo
  {author} {\bibfnamefont {M.}~\bibnamefont {Singh}},\ }\href@noop {}
  {\bibfield  {journal} {\bibinfo  {journal} {Physical Review E}\ }\textbf
  {\bibinfo {volume} {97}},\ \bibinfo {pages} {063003} (\bibinfo {year}
  {2018})}\BibitemShut {NoStop}%
\bibitem [{\citenamefont {Parisi}\ \emph {et~al.}(2019)\citenamefont {Parisi},
  \citenamefont {Procaccia}, \citenamefont {Shor},\ and\ \citenamefont
  {Zylberg}}]{19PPSZ}%
  \BibitemOpen
  \bibfield  {author} {\bibinfo {author} {\bibfnamefont {G.}~\bibnamefont
  {Parisi}}, \bibinfo {author} {\bibfnamefont {I.}~\bibnamefont {Procaccia}},
  \bibinfo {author} {\bibfnamefont {C.}~\bibnamefont {Shor}}, \ and\ \bibinfo
  {author} {\bibfnamefont {J.}~\bibnamefont {Zylberg}},\ }\href@noop {}
  {\bibfield  {journal} {\bibinfo  {journal} {Physical Review E}\ }\textbf
  {\bibinfo {volume} {99}},\ \bibinfo {pages} {011001} (\bibinfo {year}
  {2019})}\BibitemShut {NoStop}%
\bibitem [{\citenamefont {Born}\ and\ \citenamefont {Huang}(1998)}]{Born98}%
  \BibitemOpen
  \bibfield  {author} {\bibinfo {author} {\bibfnamefont {M.}~\bibnamefont
  {Born}}\ and\ \bibinfo {author} {\bibfnamefont {K.}~\bibnamefont {Huang}},\
  }\href@noop {} {\emph {\bibinfo {title} {Dynamical theory of crystal
  lattices}}}\ (\bibinfo  {publisher} {Oxford university press},\ \bibinfo
  {year} {1998})\BibitemShut {NoStop}%
\bibitem [{\citenamefont {Maradudin}\ \emph {et~al.}(1963)\citenamefont
  {Maradudin}, \citenamefont {Montroll}, \citenamefont {Weiss},\ and\
  \citenamefont {Ipatova}}]{63MMWW}%
  \BibitemOpen
  \bibfield  {author} {\bibinfo {author} {\bibfnamefont {A.~A.}\ \bibnamefont
  {Maradudin}}, \bibinfo {author} {\bibfnamefont {E.~W.}\ \bibnamefont
  {Montroll}}, \bibinfo {author} {\bibfnamefont {G.~H.}\ \bibnamefont {Weiss}},
  \ and\ \bibinfo {author} {\bibfnamefont {I.}~\bibnamefont {Ipatova}},\
  }\href@noop {} {\emph {\bibinfo {title} {Theory of lattice dynamics in the
  harmonic approximation}}},\ Vol.~\bibinfo {volume} {3}\ (\bibinfo
  {publisher} {Academic press New York},\ \bibinfo {year} {1963})\BibitemShut
  {NoStop}%
\bibitem [{\citenamefont {Squire}\ and\ \citenamefont {Hoover}(1969)}]{69SHH}%
  \BibitemOpen
  \bibfield  {author} {\bibinfo {author} {\bibfnamefont {A.}~\bibnamefont
  {Squire}, \bibfnamefont {D.R.~Holt}}\ and\ \bibinfo {author} {\bibfnamefont
  {W.}~\bibnamefont {Hoover}},\ }\href {\doibase 10.1016/0031-8914(69)90031-7}
  {\bibfield  {journal} {\bibinfo  {journal} {Physica}\ }\textbf {\bibinfo
  {volume} {42}},\ \bibinfo {pages} {388} (\bibinfo {year} {1969})}\BibitemShut
  {NoStop}%
\bibitem [{\citenamefont {Lutsko}(1989)}]{89Lut}%
  \BibitemOpen
  \bibfield  {author} {\bibinfo {author} {\bibfnamefont {J.}~\bibnamefont
  {Lutsko}},\ }\href@noop {} {\bibfield  {journal} {\bibinfo  {journal}
  {Journal of applied physics}\ }\textbf {\bibinfo {volume} {65}},\ \bibinfo
  {pages} {2991} (\bibinfo {year} {1989})}\BibitemShut {NoStop}%
\bibitem [{\citenamefont {Stillinger}\ and\ \citenamefont
  {Weber}(1984)}]{84SW}%
  \BibitemOpen
  \bibfield  {author} {\bibinfo {author} {\bibfnamefont {F.~H.}\ \bibnamefont
  {Stillinger}}\ and\ \bibinfo {author} {\bibfnamefont {T.~A.}\ \bibnamefont
  {Weber}},\ }\href@noop {} {\bibfield  {journal} {\bibinfo  {journal} {The
  Journal of chemical physics}\ }\textbf {\bibinfo {volume} {80}},\ \bibinfo
  {pages} {4434} (\bibinfo {year} {1984})}\BibitemShut {NoStop}%
\bibitem [{\citenamefont {Maloney}\ and\ \citenamefont
  {Lema\^{i}tre}(2006)}]{06MLa}%
  \BibitemOpen
  \bibfield  {author} {\bibinfo {author} {\bibfnamefont {C.}~\bibnamefont
  {Maloney}}\ and\ \bibinfo {author} {\bibfnamefont {A.}~\bibnamefont
  {Lema\^{i}tre}},\ }\href {\doibase 10.1103/PhysRevE.74.016118} {\bibfield
  {journal} {\bibinfo  {journal} {Phys. Rev. E}\ }\textbf {\bibinfo {volume}
  {74}},\ \bibinfo {pages} {016118} (\bibinfo {year} {2006})}\BibitemShut
  {NoStop}%
\bibitem [{\citenamefont {Hentschel}\ \emph {et~al.}(2011)\citenamefont
  {Hentschel}, \citenamefont {Karmakar}, \citenamefont {Lerner},\ and\
  \citenamefont {Procaccia}}]{11HKLP}%
  \BibitemOpen
  \bibfield  {author} {\bibinfo {author} {\bibfnamefont {H.~G.~E.}\
  \bibnamefont {Hentschel}}, \bibinfo {author} {\bibfnamefont {S.}~\bibnamefont
  {Karmakar}}, \bibinfo {author} {\bibfnamefont {E.}~\bibnamefont {Lerner}}, \
  and\ \bibinfo {author} {\bibfnamefont {I.}~\bibnamefont {Procaccia}},\ }\href
  {\doibase 10.1103/PhysRevE.83.061101} {\bibfield  {journal} {\bibinfo
  {journal} {Phys. Rev.E}\ }\textbf {\bibinfo {volume} {83}},\ \bibinfo {pages}
  {061101} (\bibinfo {year} {2011})}\BibitemShut {NoStop}%
\bibitem [{\citenamefont {Dasgupta}\ \emph
  {et~al.}(2012{\natexlab{a}})\citenamefont {Dasgupta}, \citenamefont
  {Karmakar},\ and\ \citenamefont {Procaccia}}]{12DKP}%
  \BibitemOpen
  \bibfield  {author} {\bibinfo {author} {\bibfnamefont {R.}~\bibnamefont
  {Dasgupta}}, \bibinfo {author} {\bibfnamefont {S.}~\bibnamefont {Karmakar}},
  \ and\ \bibinfo {author} {\bibfnamefont {I.}~\bibnamefont {Procaccia}},\
  }\href {\doibase 10.1103/PhysRevLett.108.075701} {\bibfield  {journal}
  {\bibinfo  {journal} {Phys. Rev. Lett.}\ }\textbf {\bibinfo {volume} {108}},\
  \bibinfo {pages} {075701} (\bibinfo {year} {2012}{\natexlab{a}})},\ \Eprint
  {http://arxiv.org/abs/1111.0823} {arXiv:1111.0823} \BibitemShut {NoStop}%
\bibitem [{\citenamefont {Dasgupta}\ \emph
  {et~al.}(2012{\natexlab{b}})\citenamefont {Dasgupta}, \citenamefont
  {Hentschel},\ and\ \citenamefont {Procaccia}}]{12DHP}%
  \BibitemOpen
  \bibfield  {author} {\bibinfo {author} {\bibfnamefont {R.}~\bibnamefont
  {Dasgupta}}, \bibinfo {author} {\bibfnamefont {H.~G.~E.}\ \bibnamefont
  {Hentschel}}, \ and\ \bibinfo {author} {\bibfnamefont {I.}~\bibnamefont
  {Procaccia}},\ }\href
  {https://link.aps.org/doi/10.1103/PhysRevLett.109.255502} {\bibfield
  {journal} {\bibinfo  {journal} {Phys. Rev. Lett.}\ }\textbf {\bibinfo
  {volume} {109}},\ \bibinfo {pages} {255502} (\bibinfo {year}
  {2012}{\natexlab{b}})}\BibitemShut {NoStop}%
\bibitem [{\citenamefont {Ghosh}\ \emph {et~al.}(2010)\citenamefont {Ghosh},
  \citenamefont {Chikkadi}, \citenamefont {Schall}, \citenamefont {Kurchan},\
  and\ \citenamefont {Bonn}}]{10GCDKB}%
  \BibitemOpen
  \bibfield  {author} {\bibinfo {author} {\bibfnamefont {A.}~\bibnamefont
  {Ghosh}}, \bibinfo {author} {\bibfnamefont {V.~K.}\ \bibnamefont {Chikkadi}},
  \bibinfo {author} {\bibfnamefont {P.}~\bibnamefont {Schall}}, \bibinfo
  {author} {\bibfnamefont {J.}~\bibnamefont {Kurchan}}, \ and\ \bibinfo
  {author} {\bibfnamefont {D.}~\bibnamefont {Bonn}},\ }\href {\doibase
  10.1103/PhysRevLett.104.248305} {\bibfield  {journal} {\bibinfo  {journal}
  {Phys. Rev. Lett.}\ }\textbf {\bibinfo {volume} {104}},\ \bibinfo {pages}
  {248305} (\bibinfo {year} {2010})}\BibitemShut {NoStop}%
\bibitem [{\citenamefont {Henkes}\ \emph {et~al.}(2012)\citenamefont {Henkes},
  \citenamefont {Brito},\ and\ \citenamefont {Dauchot}}]{12HBD}%
  \BibitemOpen
  \bibfield  {author} {\bibinfo {author} {\bibfnamefont {S.}~\bibnamefont
  {Henkes}}, \bibinfo {author} {\bibfnamefont {C.}~\bibnamefont {Brito}}, \
  and\ \bibinfo {author} {\bibfnamefont {O.}~\bibnamefont {Dauchot}},\ }\href
  {\doibase 10.1039/C2SM07445A} {\bibfield  {journal} {\bibinfo  {journal}
  {Soft Matter}\ }\textbf {\bibinfo {volume} {8}},\ \bibinfo {pages} {6092}
  (\bibinfo {year} {2012})}\BibitemShut {NoStop}%
\bibitem [{\citenamefont {Schindler}\ and\ \citenamefont {Maggs}(2016)}]{16SM}%
  \BibitemOpen
  \bibfield  {author} {\bibinfo {author} {\bibfnamefont {M.}~\bibnamefont
  {Schindler}}\ and\ \bibinfo {author} {\bibfnamefont {A.}~\bibnamefont
  {Maggs}},\ }\href@noop {} {\bibfield  {journal} {\bibinfo  {journal} {Soft
  matter}\ }\textbf {\bibinfo {volume} {12}},\ \bibinfo {pages} {2612}
  (\bibinfo {year} {2016})}\BibitemShut {NoStop}%
\bibitem [{\citenamefont {Perera}\ and\ \citenamefont
  {Harrowell}(1999)}]{99PH}%
  \BibitemOpen
  \bibfield  {author} {\bibinfo {author} {\bibfnamefont {D.~N.}\ \bibnamefont
  {Perera}}\ and\ \bibinfo {author} {\bibfnamefont {P.}~\bibnamefont
  {Harrowell}},\ }\href {\doibase 10.1103/PhysRevE.59.5721} {\bibfield
  {journal} {\bibinfo  {journal} {Phys. Rev. E}\ }\textbf {\bibinfo {volume}
  {59}},\ \bibinfo {pages} {5721} (\bibinfo {year} {1999})}\BibitemShut
  {NoStop}%
\bibitem [{\citenamefont {Kob}\ and\ \citenamefont {Andersen}(1994)}]{94KA}%
  \BibitemOpen
  \bibfield  {author} {\bibinfo {author} {\bibfnamefont {W.}~\bibnamefont
  {Kob}}\ and\ \bibinfo {author} {\bibfnamefont {H.~C.}\ \bibnamefont
  {Andersen}},\ }\href {\doibase 10.1103/PhysRevLett.73.1376} {\bibfield
  {journal} {\bibinfo  {journal} {Phys. Rev. Lett.}\ }\textbf {\bibinfo
  {volume} {73}},\ \bibinfo {pages} {1376} (\bibinfo {year}
  {1994})}\BibitemShut {NoStop}%
\end{thebibliography}%

\end{document}